\documentclass[journal, twoside]{IEEEtran}

\usepackage{cite}
\usepackage{xcolor}
\usepackage{amsmath}
\usepackage{amssymb}
\usepackage{amsfonts}
\usepackage{graphicx}
\usepackage{textcomp, dirtytalk}
\usepackage{algorithm, algorithmic}
\newcommand{\revision}[1]{#1}
\usepackage[hidelinks]{hyperref}
\DeclareMathOperator{\Min}{Min}

\begin{document}

\title{MultPIM: Fast Stateful Multiplication for Processing-in-Memory}

\markboth{Accepted To IEEE Transactions On Circuits And Systems—II: Express Briefs}{Leitersdorf \MakeLowercase{\textit{et al.}}: Fast Stateful Multiplication for Processing-in-Memory}

\author{Orian~Leitersdorf,~\IEEEmembership{Member,~IEEE,}
        Ronny~Ronen,~\IEEEmembership{Fellow,~IEEE,}
        and~Shahar~Kvatinsky,~\IEEEmembership{Senior~Member,~IEEE}%
\thanks{Manuscript received June 23, 2021; revised September 6, 2021; accepted September 20, 2021. This work was supported in part by the European Research Council through the European Union's Horizon 2020 Research and Innovation Programe under Grant 757259, and in part by the Israel Science Foundation under Grant 1514/17. (\textit{Corresponding author: Orian Leitersdorf.})
}%
\thanks{Orian Leitersdorf, Ronny Ronen, and Shahar Kvatinsky are with the Technion - Israel Institute of Technology, Haifa 3200003, Israel (e-mail: orianl@campus.technion.ac.il; shahar@ee.technion.ac.il).\vspace{-20pt}}%
}

\IEEEoverridecommandlockouts
\IEEEaftertitletext{\vspace{-30pt}}

\IEEEpubid{\begin{minipage}{\textwidth}\ \\[12pt] \centering \copyright 2021 IEEE. Personal use of this material is permitted.  Permission from IEEE must be obtained for all other uses, in any current or future media, including reprinting/republishing this material for advertising or promotional purposes, creating new collective works, for resale or redistribution to servers or lists, or reuse of any copyrighted component of this work in other works.
\end{minipage}} 

\maketitle

\begin{abstract}
Processing-in-memory (PIM) seeks to eliminate computation/memory data transfer using devices that support both storage and logic. Stateful logic techniques such as IMPLY, MAGIC and FELIX can perform logic gates within memristive crossbar arrays with massive parallelism. Multiplication via stateful logic is an active field of research due to the wide implications. Recently, RIME has become the state-of-the-art algorithm for stateful single-row multiplication by using memristive partitions, reducing the latency of the previous state-of-the-art by $\mbox{\boldmath $5.1\times$}$. In this paper, we begin by proposing novel partition-based computation techniques for broadcasting and shifting data. Then, we design an in-memory multiplication algorithm based on the carry-save add-shift (CSAS) technique. Finally, we \revision{develop a novel stateful full-adder that significantly improves the state-of-the-art (FELIX) design}. These contributions constitute MultPIM, a multiplier that reduces state-of-the-art time complexity from quadratic to linear-log. For 32-bit numbers, MultPIM improves latency by an \emph{additional} \revision{$\mbox{\boldmath $4.2\times$}$} over RIME, while even slightly reducing area overhead. Furthermore, we optimize MultPIM for full-precision matrix-vector multiplication and improve latency by \revision{$\mbox{\boldmath $25.5\times$}$} over FloatPIM matrix-vector multiplication.
\end{abstract}

\vspace{-5pt}


\vspace{-12pt}

\IEEEpubidadjcol

\section{Introduction}
\label{sec:introduction}

\IEEEPARstart{T}{he} von Neumann architecture separates computation and memory in computing systems. Each has significantly improved in recent years, leading to the data-transfer between them becoming a bottleneck (\emph{memory wall}\cite{DarkMemory}). \emph{Processing-in-Memory} (PIM) aims to nearly eliminate this data-transfer by using devices that support both storage and logic.

Processing-in-memory can be implemented using \emph{memristors}~\cite{Memristor}, two-terminal devices with variable resistance. Their resistance may represent binary data by being set to either low-resistive state (LRS) or high-resistive state (HRS). A high-density memory can be built using a memristor crossbar array structure~\cite{DesiredMemristor}. Uniquely, the resistance of a memristor can be controlled via an applied voltage, enabling stateful logic to be performed within the crossbar array. \revision{While there remain various challenges with memristive memory and stateful logic, promising ongoing research has experimentally demonstrated stateful logic~\cite{BarakVCM, LogicComputing} and proposed solutions for reliable operation~\cite{InSituAging, EfficientTesting, ReliableNonVoltatileMemory}. Therefore, we assume the widely-accepted stateful-logic model~\cite{MemristiveLogic} and focus on algorithmic aspects.} Examples of stateful logic techniques include IMPLY~\cite{borghetti2010memristive}, MAGIC~\cite{MAGIC} and FELIX~\cite{FELIX}, which can also be performed in parallel along rows/columns. Hence, single-row computation algorithms are advantageous as they can be repeated along all rows with the exact same latency. Additional parallelism can arise from memristive partitions~\cite{FELIX} which dynamically divide the crossbar array using transistors. In this paper, we propose novel partition-based techniques for efficiently broadcasting/shifting data amongst partitions.

\IEEEpubidadjcol

Multiplication is fundamental for many applications, \textit{e.g.}, convolution and matrix-multiplication. Initially, only non-crossbar-compatible and non-single-row algorithms~\cite{Skeleton, UltraEfficient, Dadda, SemiSerialMultiplier, MemristorBasedMultipliers, ResistiveComputing} for in-memory multiplication were considered. Yet, these algorithms only support multiplying two numbers per \emph{crossbar}, rather than per \emph{crossbar row} -- which would enable paralleled element-wise vector multiplication. The first in-row multiplication algorithm was proposed by Haj-Ali \textit{et al.}~\cite{Ameer}, and was later utilized in IMAGING~\cite{IMAGING} for image processing and in FloatPIM~\cite{FloatPIM} for deep neural networks. This algorithm requires $O(N^2)$ latency and $O(N)$ memristors, where $N$ is the width of each number. Recently, RIME~\cite{RIME} improved the latency by $5.1 \times$ for $N=32$ via memristive partitions~\cite{FELIX}, while slightly reducing area (\textit{i.e.} memristor count) as well. The asymptotic latency/area remains at $O(N^2)$ and $O(N)$ (respectively). RIME is based on Wallace tree computation using $N-1$ partitions in a single row, each partition representing a full-adder unit. The bottleneck of RIME is the partial product computation and data-transfer between partitions (as they occur serially), accounting for $81\%$ of the latency. 

In this paper, we speedup multiplication using three methods. First, we propose novel partition-based computation techniques for broadcasting/shifting data amongst partitions. Second, we replace the Wallace tree with a carry-save-add-shift (CSAS) multiplier~\cite{sunder1995two, richards1955arithmetic, FSP}. Lastly, \revision{we propose a novel full-adder design that significantly improves the previous state-of-the-art (FELIX~\cite{FELIX}).} The final algorithm, coined MultPIM, achieves an asymptotic latency of $O(N\log N)$ with $O(N)$ area. For $N=32$, MultPIM achieves a \revision{$4.2\times$} improvement in latency over RIME (that is, \revision{$21.1\times$} over Haj-Ali \textit{et al.}) while maintaining constant partition count and even slightly reducing area. This paper contributes the following:
\begin{itemize}
    \item \emph{Partition Techniques:} Introduces novel techniques for broadcasting/shifting data amongst partitions.
    \item \revision{\emph{Full Adder:} Proposes a full-adder design that improves the previous state-of-the-art (FELIX~\cite{FELIX}) by up to 33$\%$.}
    \item \emph{MultPIM:} Proposal of an efficient parallel multiplier that replaces quadratic time complexity with linear-log. We show latency improvement of \revision{$4.2\times$} and slight area reduction over the previous state-of-the-art (RIME~\cite{RIME}).
    \item \emph{Matrix-vector multiplication:} We present an \emph{optimized} implementation of MultPIM in matrix-vector multiplication that improves latency by \revision{$25.5\times$} over FloatPIM~\cite{FloatPIM}. 
\end{itemize}

\section{Background}
\label{sec:background}

\begin{figure}[!t]
\centering 
\includegraphics[width=3.4in]{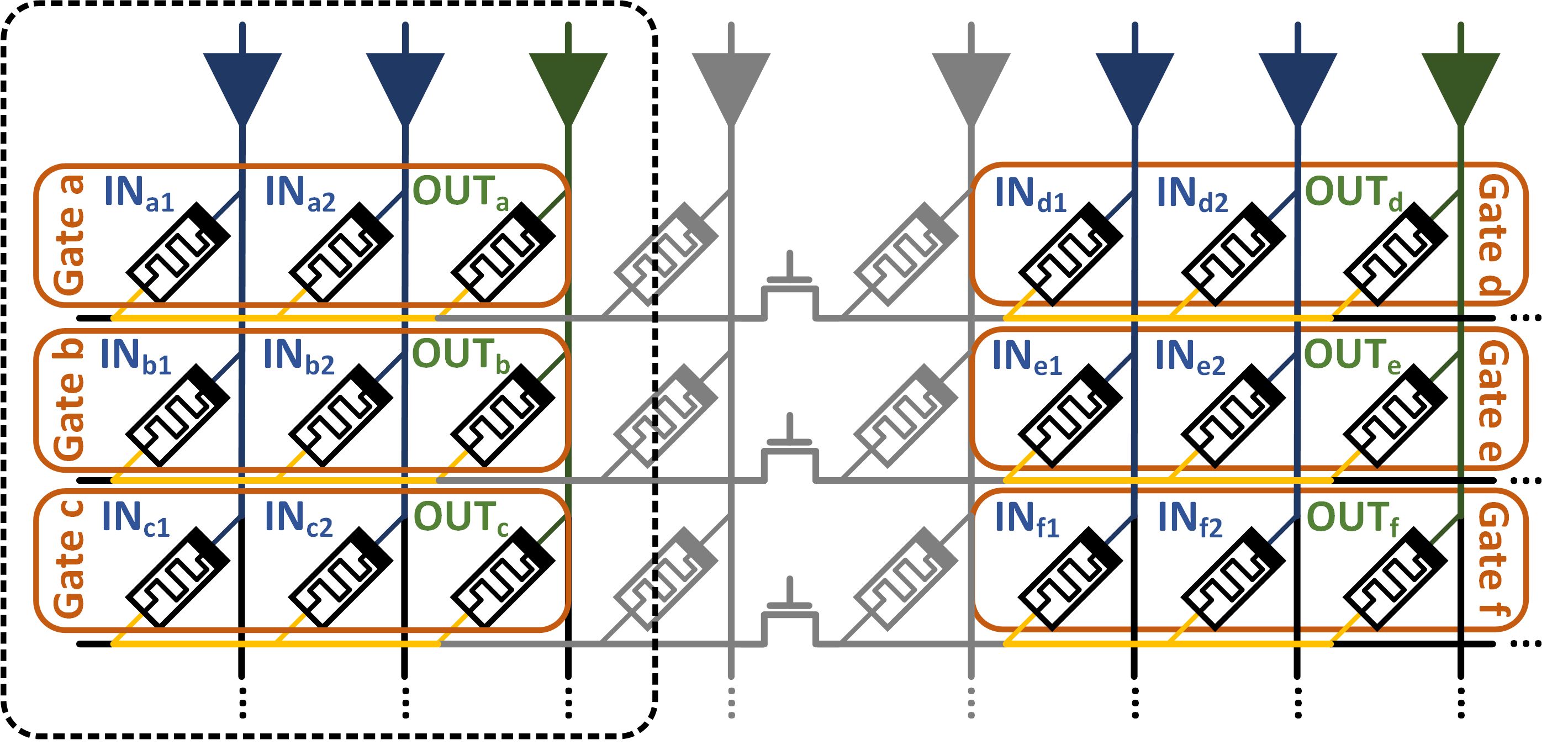}
\caption{Dashed: Memristive crossbar array with simultaneous in-row MAGIC NOR operations. Column partitions~\cite{FELIX} increase parallelism, \textit{e.g.}, performing all of the highlighted MAGIC gates in a single clock cycle.}
\label{fig:crossbar} 
\vspace{-10pt}
\end{figure}

\subsection{Stateful Logic}
\label{sec:background:stateful}

Memristive crossbar arrays have horizontal wordlines, vertical bitlines, and memristors at crosspoints. Stateful logic employs the same memristors for logic. This is possible by exploiting the unique property of memristors (voltage-controlled variable resistance). IMPLY~\cite{borghetti2010memristive}, MAGIC~\cite{MAGIC} and FELIX~\cite{FELIX} are such techniques, computing logic gates by applying voltages along either bitlines or wordlines. Together, they support logic gates such as NOT, NOR, OR, NAND, and Minority3\footnote{\revision{Haj-Ali \textit{et al.}~\cite{Ameer} assumes NOT/NOR, RIME~\cite{RIME} assumes NOT/NOR/NAND/Min3 and MultPIM assumes NOT/Min3 (fair comparison to RIME). MultPIM with other gates is included on the \href{https://github.com/oleitersdorf/MultPIM}{repository}.}}. Further, an AND with the previous output cell value can performed by skipping initialization~\cite{FELIX, XMAGIC}.

Stateful logic support massive-parallelism. The same in-row logic gate can be repeated along rows while still being performed in a single clock cycle, as seen in the dashed portion of Figure~\ref{fig:crossbar}. Essentially, in a single cycle we can perform a single element-wise logic operation on columns of a crossbar. Hence, memristive computation algorithms are typically limited to a single row of memristors as this allows repetition of the algorithm along many rows (\textit{e.g.}, for vector operation) in the same latency~\cite{SIMPLER}. This parallelism can be further increased through memristive partitions~\cite{FELIX}. These transistors divide the crossbar into \emph{partitions} and can be dynamically set to either non-conducting (for parallel operation amongst the partitions, see Figure~\ref{fig:crossbar}) or conducting (for logic between partitions).

\vspace{-8pt}

\subsection{Carry-Save Add-Shift (CSAS)}
\label{sec:background:csas}

\begin{figure}[!t]
\centering 
\includegraphics[width=\linewidth]{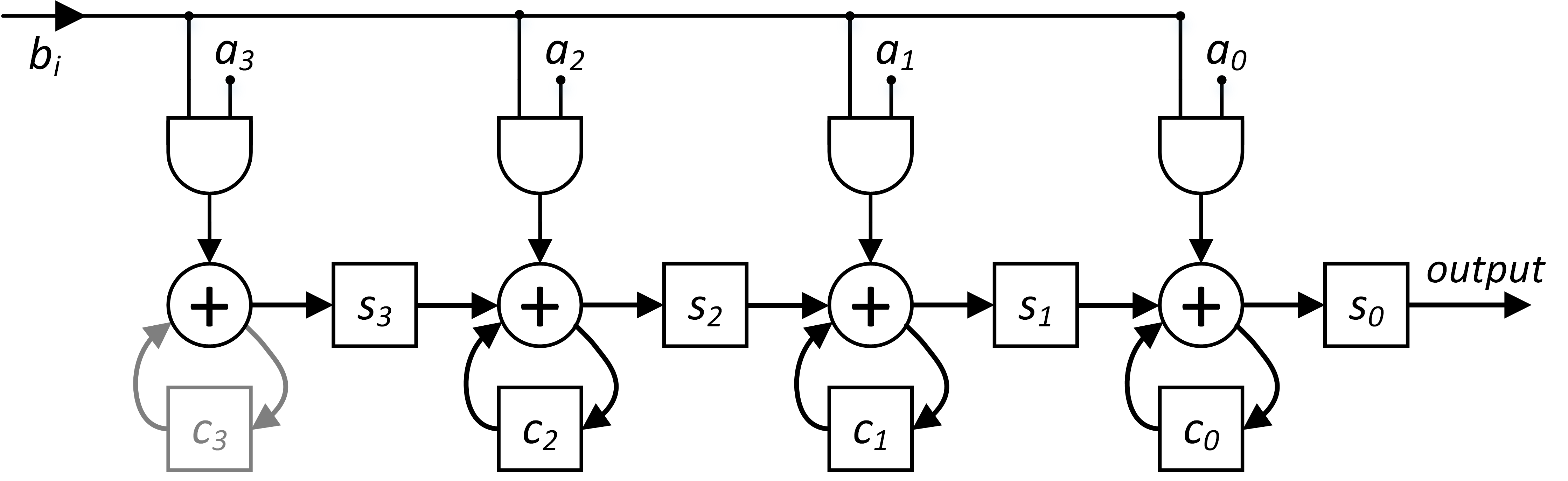}
\caption{Four-bit CSAS multiplier~\cite{FSP}. In each cycle, a single bit from input $b$ is fed to compute that corresponding partial product; a carry-save adder adds this partial-product to the current sum/carry. Notice that $c_3$ is always zero. Latches are squares and full-adders are circles.
}
\label{fig:csas}
\vspace{-10pt}
\end{figure}

The carry-save add-shift (CSAS) technique~\cite{sunder1995two, richards1955arithmetic, FSP} utilizes a carry-save adder~\cite{ComputerArithmetic} for multiplication. The technique stores two numbers, the current sum and the current carry, and adds the partial products to these numbers using the carry-save adder. This can be more efficient than a traditional \emph{shift-and-add} multiplier as carry propagation is avoided in intermediate steps. Rather than shift the partial products, the CSAS technique shifts the sum -- effectively emulating moving full adders (FAs). Figure~\ref{fig:csas} details the overall circuit. In $N$ stages, this circuit produces the lower $N$ bits of the product. The top $N$ bits can be computed as the sum of the final sum and carry numbers. These two numbers can be added by feeding zero partial products to the FAs for $N$ stages, or with a regular adder (\textit{e.g.}, ripple carry)~\cite{sunder1995two, richards1955arithmetic, FSP}. 

\begin{figure*}
\centering 
\includegraphics[width=6.8in]{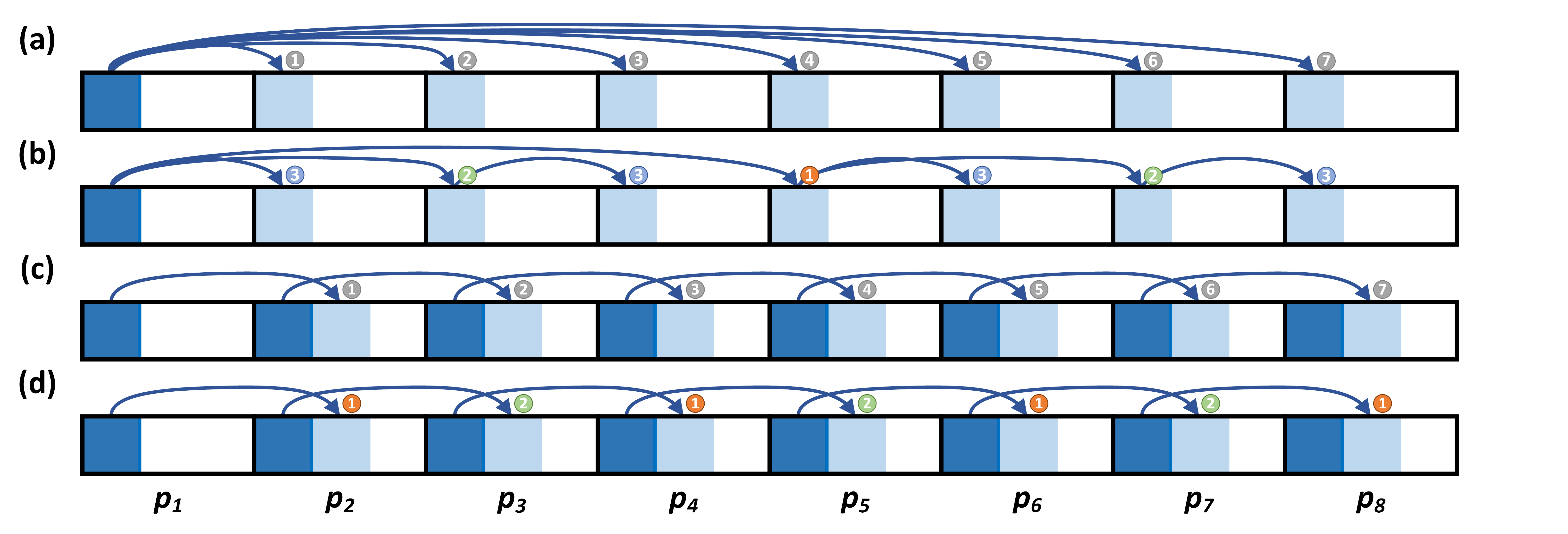}
\caption{(a) The naive solution to the broadcasting task, requiring $k-1$ cycles, and (b) the proposed solution requiring $\log_2 k$ cycles. (c) The naive solution to the shift task, requiring $k-1$ cycles, and (d) the proposed solution requiring $2$ cycles. Circled numbers represent the cycle number.}
\label{fig:techniques} 
\vspace{-5pt}
\end{figure*}

\section{Partition Techniques}
\label{sec:partitionTechniques}
In this section, we introduce two novel partition techniques. The first technique involves \emph{broadcasting} a single bit from one partition to $k$ partitions in $\log_2(k)$ cycles, and the second technique involves \emph{shifting} bits across $k$ partitions in two cycles. Throughout this section, we assume $k$ consecutive partitions and $p_i$ refers to the $i^{th}$ partition. 

For simplicity, we do not discuss initialization cycles in this section and we also assume the existence of a \emph{copy} gate: similar to MAGIC NOT, but without negation. Note that the final MultPIM implementation accounts for initialization cycles and does not require a \emph{copy} gate.

\vspace{-5pt}

\subsection{Broadcasting Technique}
\label{sec:partitionTechniques:oneToAll}
Assume that partition $p_1$ contains a bit that we want to transfer to all of the other partitions. The naive approach illustrated in Figure~\ref{fig:techniques}(a) will perform the operation serially: copying the bit from the first partition to each of the others, one at a time, for a total of $k-1$ clock cycles. In terms of area, this naive approach requires one memristor from each partition (no extra intermediate memristors are necessary).

We propose a novel recursive technique for solving this task, dynamically selecting the partition transistors. We begin by copying the bit from $p_1$ to $p_{k/2+1}$. Then, we set the transistor between $p_{k/2}$ and $p_{k/2+1}$ to non-conducting and proceed recursively \emph{in parallel} with $p_1, ..., p_{k/2}$ and $p_{k/2+1}, ..., p_k$. This technique requires a total of $\log_2k$ cycles and only one memristor per partition (no extra intermediate memristors necessary), as shown in Figure~\ref{fig:techniques}(b). 

\vspace{-5pt}

\subsection{Shift Technique}
\label{sec:partitionTechniques:shift}
Assume that each partition begins with its own bit, and that we want to shift these bits between the partitions: the bit from $p_1$ moves to $p_2$, the bit from $p_2$ moves to $p_3$, ..., the bit from $p_{k-1}$ moves to $p_k$. RIME performs this transfer in $k-1$ cycles as shown in Figure~\ref{fig:techniques}(c). In terms of area, this technique requires no additional intermediate memristors.

We propose a novel technique involving only two steps: copying from all odd partitions to even partitions, and then copying from all even partitions to odd partitions. This technique is demonstrated in Figure~\ref{fig:techniques}(d), utilizing exactly $2$ clock cycles in total. Note that we can replace the copy gate with any other logic gate (\textit{i.e.}, storing multiple input bits in each partition, and storing the output of the logic gate on the inputs of the $i^{th}$ partition in the $i+1^{th}$ partition). This concept is utilized in Section~\ref{sec:fsm:implementation} to optimize full-adder logic. 

\vspace{-10pt}

\section{MultPIM: Fast Stateful Multiplier}
\label{sec:fsm}
In this section, we combine the CSAS multiplier with the two novel techniques from Section~\ref{sec:partitionTechniques} to introduce MultPIM. We begin by describing the general algorithm concept, and then continue by providing various optimizations for latency and area. Throughout this section, let $a = (a_{N-1}...a_0)_2$ and $b = (b_{N-1}...b_0)_2$\footnote{MultPIM also supports different widths for $a$ and $b$.}; we are interested in computing $a\cdot b$. Recall that $p_i$ is the $i^{th}$ partition, and let $p_i.x$ represent the variable $x$ stored in $p_i$ (single bit).

\begin{figure*}[!t]
\centering 
\includegraphics[width=6.5in]{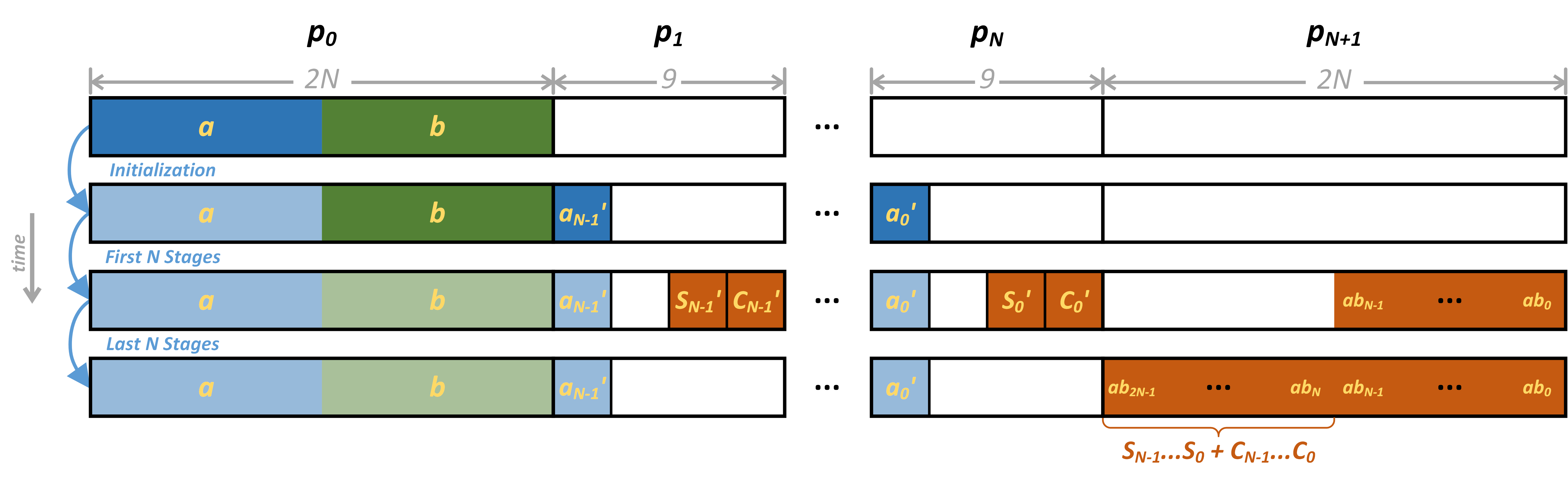}
\caption{The main steps of the MultPIM algorithm. Note that the last $N$ bits of the product are the sum of $S_{N-1}...S_0$ and $C_{N-1}...C_0$; this sum can either be computed via the \emph{Last $N$ Stages}, or by using a conventional adder. Faded-out cells indicate values no longer used.}
\label{fig:fsm} 
\vspace{-10pt}
\end{figure*}

\vspace{-8pt}

\subsection{General Algorithmic Concept}
\label{sec:fsm:concept}
The general concept involves using $N$ full-adders in parallel (each in a partition), similar to the CSAS technique (see Figure~\ref{fig:csas}). Each partition stores one bit of $a$ throughout the entire computation (\textit{i.e.}, partition $p_i$ stores $a_{N-i}$). In addition, each partition stores carry/sum bits (similar to CSAS latches).

Following the CSAS technique, the computation begins with $N$ stages in which $b$ is fed into the system. For the $i^{th}$ stage in the first $N$ stages, we perform the following:
\begin{itemize}
    \item Copy $b_i$ to all of the partitions using the technique from Section~\ref{sec:partitionTechniques:oneToAll} in $\log_2 N$ cycles. 
    \item Compute the partial product in all of the partitions in parallel (similar to AND gates in CSAS).
    \item Compute full-adder in each of the partitions in parallel, using the stored carry/sum bits and the partial product bit. The new sum/carry replace the old sum/carry bits.
    \item Shift the sum bits amongst the partitions using the technique from Section~\ref{sec:partitionTechniques:shift}. Lowest bit is stored as output.
\end{itemize}

We choose\footnote{A regular adder can be implemented instead in $p_{N+1}$. During that time, partitions $p_0, p_1, ..., p_N$ could compute the product of a different independent pair of numbers as part of a multiplication pipeline.} to proceed by feeding another $N$ zeros for $b$ to propagate the stored carries. That is, the algorithm continues with another $N$ stages as follows:
\begin{itemize}
    \item Compute half-adder in each of the partitions in parallel, using the stored carry-bit and the stored sum-bit. The new sum/carry bits replace the old ones.
    \item Shift the sum bits amongst the partitions using the technique from Section~\ref{sec:partitionTechniques:shift}. Lowest bit is stored as output.
\end{itemize}

Overall, $N$ stages with latency $O(\log_2N)$ and another $N$ stages with latency $O(1)$. Hence, total latency is $O(N\log_2N)$. Each partition requires $O(1)$ memristors, so we require $O(N)$ memristors in total. The above stages are shown in Figure~\ref{fig:fsm}.

\vspace{-8pt}

\subsection{Implementation and Optimizations}
\label{sec:fsm:implementation}

Algorithm~\ref{alg:fsm} details the steps of the computation. Note that the usage of $\forall i$ in the \emph{\say{In parallel}} lines indicates that the computation is performed in parallel on all partitions. The for loops in the algorithm are evaluated serially. We detail here various specific optimizations to the algorithm.

\subsubsection{\revision{Full Adder}} 
\revision{
The state-of-the-art\footnote{\revision{The full-adder proposed by RIME~\cite{RIME} requires 7 cycles. Note that our novel full-adder is \emph{inspired} by the expressions from RIME.}} (FELIX~\cite{FELIX}) requires 6 cycles (without init.), assumes NOT/OR/NAND/Min3, and requires 2 intermediates. Our novel full-adder is based on:}
\begin{equation}
    \revision{C_{out} = \Min_3'(A, B, C_{in}),}
\end{equation}
\begin{equation}
    \revision{S_{out} = \Min_3(C_{out}, C_{in}', \Min_3(A, B, C_{in}')).}
\end{equation}
\revision{The improvement over FELIX~\cite{FELIX} originates from using $C_{out}$ for computing $S$. These expressions enable $5$ cycles, assuming only NOT/Min3, and requiring 3 intermediate memristors\footnote{\revision{6 cycles, assuming NOT/Min3, and only 2 intermediate memristors is possible with re-use. Therefore, FELIX~\cite{FELIX} is replaced completely.}}. Further, if the not of an input is also given, only $4$ cycles are required (\textit{i.e.}, no need to compute $C_{in}'$)\footnote{\revision{This enables $N$-bit addition with $5N$ cycles and $3N+5$ memristors using only NOT/Min3, compared to $7N$ and $3N+2$ from FELIX (including init.).}}. The latter is utilized for Lines~\ref{alg:fsm:firstFor:fa}-\ref{alg:fsm:firstFor:shift} and Lines~\ref{alg:fsm:lastFor:ha}-\ref{alg:fsm:lastFor:shift} by storing both $C, C'$ and performing the sum computation as part of shift.}

\subsubsection{Lines~\ref{alg:fsm:firstFor:b}-\ref{alg:fsm:firstFor:ab}} Performing the Section~\ref{sec:partitionTechniques:oneToAll} algorithm with NOT (rather than the theoretical \emph{copy}) results in some partitions receiving $b_k$ and others receiving $b_k'$. The partitions that receive $b_k$ perform Line~\ref{alg:fsm:firstFor:ab} using no-initialization NOT (see Section~\ref{sec:background:stateful}) of the stored $a_i'$ into $b_k$, resulting in $(a_i')' \cdot b_k = a_i \cdot b_k$. Those that receive $b_k'$ perform Line~\ref{alg:fsm:firstFor:ab} using $\Min_3(a_i',b_k',1) = a_i \cdot b_k$. Thus, Line~\ref{alg:fsm:firstFor:ab} requires 1 cycle.

\subsubsection{Partitions}
Note that $p_0$ and $p_{N+1}$ can be merged with $p_1$ and $p_N$ (respectively) to reach a total of $N$ partitions. Furthermore, since the top carry bit is always zero (see Figure~\ref{fig:csas}), then we can use $N-1$ partitions rather than $N$.

\begin{algorithm}[t]
 \caption{MultPIM}
 \begin{algorithmic}[1]
 \renewcommand{\algorithmicrequire}{\textbf{Input:}}
 \renewcommand{\algorithmicensure}{\textbf{Output:}}
 \REQUIRE $a, b$ stored in $p_0$ (start of the row)
 \ENSURE $a \cdot b$ stored in $p_{N+1}$ (end of the row)
 \\ \textit{Initialization:}
  \STATE $\forall i \;:\; p_i.c, p_i.s \gets 0$ \COMMENT{In parallel, init. carry/sum}
  \label{alg:fsm:SC}
  \STATE \algorithmicfor \  $i = 1$ to $N$ \algorithmicdo \  $p_i.a \gets p_0.{a_{N-i}}$ \COMMENT{Store $a_{N-i}$ in $p_i$}
  \label{alg:fsm:initFor}
 \\ \textit{First $N$ Stages:}
  \FOR {$k = 1$ to $N$}
  \label{alg:fsm:firstFor}
  \STATE $\forall i \;:\; p_i.b = b_k$ \COMMENT{Using Section~\ref{sec:partitionTechniques:oneToAll}}
  \label{alg:fsm:firstFor:b}
  \STATE $\forall i \;:\; p_i.ab = p_i.a \cdot p_i.b$ \COMMENT{In parallel}
  \label{alg:fsm:firstFor:ab}
  \STATE $\forall i \;:\; p_i.s, p_i.c = FA(p_i.s,p_i.c,p_i.ab)$ \COMMENT{In parallel}
  \label{alg:fsm:firstFor:fa}
  \STATE $\forall i \;:\; p_{i+1}.s = p_i.s$ \COMMENT{Using Section~\ref{sec:partitionTechniques:shift}}
  \label{alg:fsm:firstFor:shift}
  \ENDFOR
 \\ \textit{Last $N$ Stages:}
  \FOR {$k = 1$ to $N$}
  \label{alg:fsm:lastFor}
  \STATE $\forall i \;:\; p_i.s, p_i.c = HA(p_i.s,p_i.c)$ \COMMENT{In parallel}
  \label{alg:fsm:lastFor:ha}
  \STATE $\forall i \;:\; p_{i+1}.s = p_i.s$ \COMMENT{Using Section~\ref{sec:partitionTechniques:shift}}
  \label{alg:fsm:lastFor:shift}
  \ENDFOR 
 \end{algorithmic} 
 \label{alg:fsm}
 \end{algorithm}

\section{Results}
\label{sec:results}

We evaluate MultPIM for single-row $N$-bit multiplication. We compare MultPIM to Haj-Ali \textit{et al.} \cite{Ameer} and RIME~\cite{RIME} in terms of latency, area (memristor count), and partition count. \revision{We also present MultPIM-Area that prioritizes area over latency via additional re-use~\cite{SIMPLER}}. The results are verified by a custom cycle-accurate simulator.

\vspace{-5pt}

\subsection{Latency}
\label{sec:results:latency}

We evaluate the latency of the MultPIM algorithm in clock cycles. The algorithm begins with $N$ cycles at the start to copy $a$. Then $N$ stages which feed $b$ through the full-adders, with each stage requiring \revision{$\log_2N + 8$} cycles ($\log_2N+1$ for Lines~\ref{alg:fsm:firstFor:b}-\ref{alg:fsm:firstFor:ab}, \revision{$5$} cycles for Lines~\ref{alg:fsm:firstFor:fa}-\ref{alg:fsm:firstFor:shift}, and \revision{$1$} initialization cycle). Finally, $N$ stages at the end, each requiring \revision{$6$} cycles (\revision{$5$} for Lines~\ref{alg:fsm:lastFor:ha}-\ref{alg:fsm:lastFor:shift} and $1$ initialization cycle). Overall, \revision{$N\log_2N + 14\cdot N$} cycles. In Table~\ref{tab:latency}, we compare this latency with the previous works, demonstrating \revision{$4.2\times$} improvement over the previous state-of-the-art (RIME) for the common case of $N=32$. 

\begin{table}[t]
    \centering
    \caption{Latency (Clock Cycles)}
    \begin{tabular}{l|c|c|c}
         Algorithm & Expression & $N=16$ & $N=32$ \\
         \hline
         Haj-Ali \textit{et al.}~\cite{Ameer} & $13\cdot N^2 - 14 \cdot N + 6$ & 3110 & 12870 \\
         RIME~\cite{RIME} & $2 \cdot N^2 + 16 \cdot N -19$ & 749 & 2541 \\ 
         MultPIM & \revision{$N \cdot \log_2N + 14 \cdot N + 3$} & \revision{291} & \revision{611} \\ 
         \revision{MultPIM-Area} & \revision{$N \cdot \log_2N + 23 \cdot N + 3$} & \revision{435} & \revision{899} \\ 
    \end{tabular}
    \label{tab:latency}
    \vspace{-7pt}
\end{table}

\vspace{-10pt}

\subsection{Area}
\label{sec:results:area}

The exact number of memristors required for MultPIM is evaluated here. The computation row contains $2N$ memristors for storing the inputs, $2N$ memristors for storing the outputs, and $N$ full-adder units each requiring \revision{10} memristors total. Hence, MultPIM requires \revision{$2\cdot N + 2\cdot N + 10\cdot N = 14\cdot N$} memristors. Table~\ref{tab:area} compares this with the previous works, showing a slight improvement over the state-of-the-art (RIME). Note that MultPIM and RIME both require $N-1$ partitions\footnote{The evaluation of exact partition overhead is left for future work. Regardless, MultPIM and RIME require the same number of partitions.}.

\begin{table}[t]
    \centering
    \caption{Area (\# Memristors)}
    \begin{minipage}{\linewidth}
    \centering
    \begin{tabular}{l|c|c|c}
         Algorithm & Expression & $N=16$ & $N=32$ \\
         \hline
         Haj-Ali \textit{et al.}~\cite{Ameer} & $20\cdot N - 5$ & 315 & 635 \\
         RIME~\cite{RIME} & $15 \cdot N - 12$ & 228 & 468 \\ 
         MultPIM & \revision{$14 \cdot N - 7$} & \revision{217} & \revision{441} \\ 
         \revision{MultPIM-Area} & \revision{$10 \cdot N$} & \revision{160} & \revision{320} \\
    \end{tabular}
    \end{minipage}
    \label{tab:area}
    \vspace{-10pt}
\end{table}

\vspace{-10pt}

\subsection{Logic Simulation}
\label{sec:results:sim}
We verify the results of the algorithm with a custom cycle-accurate simulator\footnote{Available at \url{https://github.com/oleitersdorf/MultPIM}.}. The simulator models the crossbar array, and has an interface for performing \emph{operations} in-memory. The MultPIM algorithm is tested by first writing the inputs to the crossbar, then allowing MultPIM to perform in-memory \emph{operations}, and finally verifying the output. The simulator counts the exact number of \emph{operations} that MultPIM uses (including initializations), verifying the theoretical analysis.

\vspace{-3pt}

\section{Matrix-Vector Multiplication}

Here, we optimize MultPIM for matrix-vector multiplication. Formally, let $\mathbf{A}$ be an $m \times n$ matrix and let $\mathbf{x}$ be a vector of dimension $n$, we are interested in computing $\mathbf{Ax}$. Each element in the matrix/vector is a fixed-point number with $N$ bits, and the data elements are stored horizontally.

\begin{figure}[!t]
\centering 
\includegraphics[width=\linewidth]{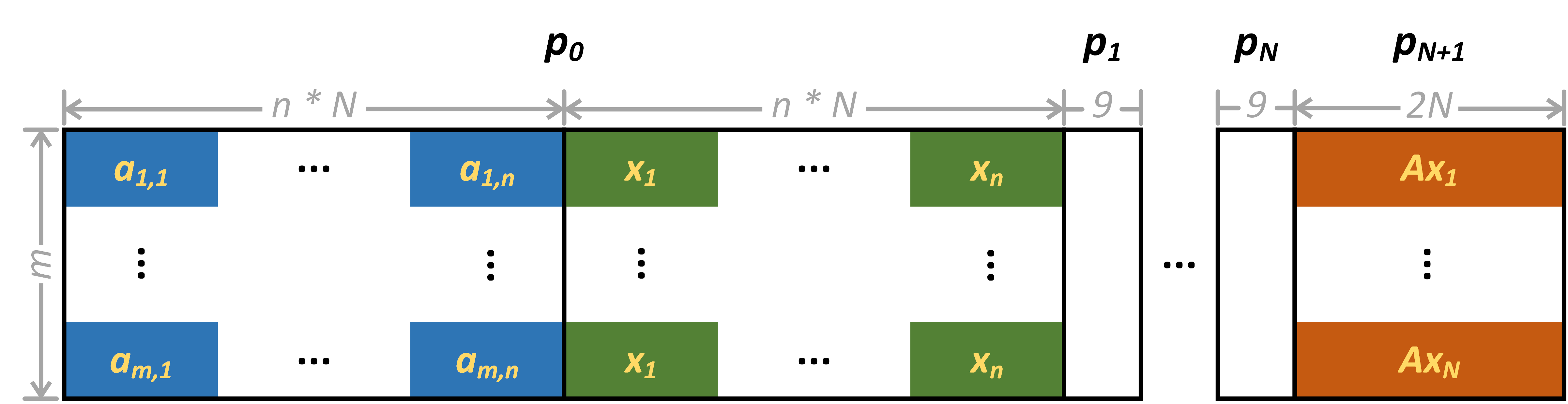}
\caption{Matrix-vector multiplication with an optimized MultPIM multiplier. Matrix $\mathbf{A}$ is shown in blue, vector $\mathbf{x}$ in green, and $\mathbf{Ax}$ in orange. The partitions are only used along columns, with the same overhead as MultPIM.}
\label{fig:maxtrixMul} 
\vspace{-10pt}
\end{figure}

The multiplication is performed by duplicating $\mathbf{x}$ along rows (see Figure~\ref{fig:maxtrixMul}), multiplying each column of the matrix with each column of the duplicated vector matrix, and then adding the results horizontally. Essentially, each row performs an inner product between the stored row of $\mathbf{A}$ and $\mathbf{x}$ (\textit{e.g.}, $\mathbf{Ax}_1 = a_{1,1}\cdot x_1 + \cdots + a_{1,n}\cdot x_n$ in the first row). A similar concept is used in FloatPIM~\cite{FloatPIM} for fixed-point matrix-multiplication. The naive solution replaces only the fixed-point multiplication algorithm in FloatPIM with MultPIM (\textit{i.e.}, compute $a_{1,1}\cdot x_1, ..., a_{1,n} \cdot x_n$ by using MultPIM $n$ times, and sum using an adder). That provides only \revision{$9.5\times$} latency improvement to FloatPIM as addition becomes non-negligible.

Instead, we optimize MultPIM to compute the sum while computing the products and further reduce product latency. The optimized algorithm receives numbers $a,b$ ($N$-bit) and $s_i,c_i$ ($2N$-bit), and computes $s_o, c_o$ ($2N$-bit) such that $s_o + c_o = a \cdot b + s_i + c_i$. This algorithm performs only \emph{Initialization} and \emph{First $N$ Stages}, thus \emph{reducing} latency compared to regular MultPIM. This is achieved by initializing the sum fields of the full-adders to the lower $N$ bits of $s_i$ (rather than zero) and feeding $p_1$ the upper bits of $s_i$ and $c_i$. The value of $s_o+c_o$ at each run of MultPIM is the sum of the products until that point. At the end, the sum $s_o + c_o$ is computed once. 

The results of the optimized matrix-vector multiplication are summarized in Table~\ref{tab:matrixMul} for $n=8,N=32$, verified by the logic simulator. \revision{We achieve $25.5\times$ latency and $1.8\times$ area improvement over FloatPIM matrix-vector multiplication, utilizing $33$ partitions. In the general case, latency is improved from $n \cdot (13 N^2 + 12 N + 6)$ to $n \cdot (N\log_2N + 11N + 9)+ 4N - 4$ cycles, and area is improved from $m \times (4 n N + 22 N - 5)$ to $m \times (2nN + 14N + 5)$ memristors, with $N+1$ partitions.}

\begin{table}[t]
    \centering
    \caption{Matrix Multiplication ($n=8, N=32$)}
    \begin{tabular}{l|c|c}
         Algorithm & Latency (Clock Cycles) & Area (Min. Crossbar Dim.) \\
         \hline
         FloatPIM & $109616$ & $m \times 1723$ \\ 
         MultPIM & \revision{$4292$} & \revision{$m \times 965$} \\
         \revision{MultPIM-Area} & \revision{$6204$} & \revision{$m \times 778$} \\
    \end{tabular}
    \label{tab:matrixMul}
    \vspace{-10pt}
\end{table}

\vspace{-5pt}

\section{Conclusion}
\label{sec:conclusion}

We present MultPIM: a novel partition-based in-memory multiplication algorithm that improves the state-of-the-art latency complexity from quadratic to linear-log, specifically by \revision{$4.2\times$} for $32$-bit. The improvement is based on the carry-save add-shift technique, two novel memristive-partition computation techniques, and \revision{an improvement to the state-of-the-art full-adder}. Furthermore, we optimize MultPIM for matrix-vector multiplication and achieve \revision{$25.5\times$}
latency and $1.8\times$ area improvements over FloatPIM matrix-vector multiplication by computing addition while performing multiplication. Correctness is verified via a cycle-accurate simulator. 

\bibliographystyle{IEEEtran}
\bibliography{refs}

\begin{thebibliography}{10}
\providecommand{\url}[1]{#1}
\csname url@samestyle\endcsname
\providecommand{\newblock}{\relax}
\providecommand{\bibinfo}[2]{#2}
\providecommand{\BIBentrySTDinterwordspacing}{\spaceskip=0pt\relax}
\providecommand{\BIBentryALTinterwordstretchfactor}{4}
\providecommand{\BIBentryALTinterwordspacing}{\spaceskip=\fontdimen2\font plus
\BIBentryALTinterwordstretchfactor\fontdimen3\font minus
  \fontdimen4\font\relax}
\providecommand{\BIBforeignlanguage}[2]{{%
\expandafter\ifx\csname l@#1\endcsname\relax
\typeout{** WARNING: IEEEtran.bst: No hyphenation pattern has been}%
\typeout{** loaded for the language `#1'. Using the pattern for}%
\typeout{** the default language instead.}%
\else
\language=\csname l@#1\endcsname
\fi
#2}}
\providecommand{\BIBdecl}{\relax}
\BIBdecl

\bibitem{DarkMemory}
A.~{Pedram}, S.~{Richardson}, M.~{Horowitz}, S.~{Galal}, and S.~{Kvatinsky},
  ``Dark memory and accelerator-rich system optimization in the dark silicon
  era,'' \emph{IEEE Design \& Test}, 2017.

\bibitem{Memristor}
L.~{Chua}, ``Memristor-the missing circuit element,'' \emph{IEEE Transactions
  on Circuit Theory}, vol.~18, no.~5, pp. 507--519, 1971.

\bibitem{DesiredMemristor}
S.~{Kvatinsky}, E.~G. {Friedman}, A.~{Kolodny}, and U.~C. {Weiser}, ``The
  desired memristor for circuit designers,'' \emph{IEEE CAS Magazine}, 2013.

\bibitem{BarakVCM}
B.~Hoffer, V.~Rana, S.~Menzel, R.~Waser, and S.~Kvatinsky, ``Experimental
  demonstration of memristor-aided logic ({MAGIC}) using valence change memory
  ({VCM}),'' \emph{IEEE Transactions on Electron Devices}, 2020.

\bibitem{LogicComputing}
Z.~Sun, E.~Ambrosi, A.~Bricalli, and D.~Ielmini, ``Logic computing with
  stateful neural networks of resistive switches,'' \emph{Advanced Materials},
  2018.

\bibitem{InSituAging}
J.~Xu, Y.~Zhan, Y.~Li, J.~Wu, X.~Ji, G.~Yu, W.~Jiang, R.~Zhao, and C.~Wang,
  ``In situ aging-aware error monitoring scheme for imply-based memristive
  computing-in-memory systems,'' \emph{IEEE TCAS-I}, 2021.

\bibitem{EfficientTesting}
P.~Liu, Z.~You, J.~Wu, B.~Liu, Y.~Han, and K.~Chakrabarty, ``Fault modeling and
  efficient testing of memristor-based memory,'' \emph{IEEE Transactions on
  Circuits and Systems I: Regular Papers}, pp. 1--12, 2021.

\bibitem{ReliableNonVoltatileMemory}
S.~Swami and K.~Mohanram, ``Reliable nonvolatile memories: Techniques and
  measures,'' \emph{IEEE Design \& Test}, 2017.

\bibitem{MemristiveLogic}
J.~Reuben, R.~Ben-Hur, N.~Wald, N.~Talati, A.~H. Ali, P.-E. Gaillardon, and
  S.~Kvatinsky, ``Memristive logic: A framework for evaluation and
  comparison,'' in \emph{PATMOS}, 2017.

\bibitem{borghetti2010memristive}
J.~Borghetti, G.~S. Snider, P.~J. Kuekes, J.~J. Yang, D.~R. Stewart, and R.~S.
  Williams, ``{‘Memristive’} switches enable ‘stateful’ logic
  operations via material implication,'' \emph{Nature}, vol. 464, no. 7290, pp.
  873--876, 2010.

\bibitem{MAGIC}
S.~Kvatinsky, D.~Belousov, S.~Liman, G.~Satat, N.~Wald, E.~G. Friedman,
  A.~Kolodny, and U.~C. Weiser, ``{MAGIC}—memristor-aided logic,'' \emph{IEEE
  Transactions on Circuits and Systems II: Express Briefs}, 2014.

\bibitem{FELIX}
S.~{Gupta}, M.~{Imani}, and T.~{Rosing}, ``{FELIX}: Fast and energy-efficient
  logic in memory,'' in \emph{ICCAD}, 2018, pp. 1--7.

\bibitem{Skeleton}
J.~Yu, R.~Nane, I.~Ashraf, M.~Taouil, S.~Hamdioui, H.~Corporaal, and
  K.~Bertels, ``Skeleton-based synthesis flow for computation-in-memory
  architectures,'' \emph{Transactions on Emerging Topics in Computing}, 2020.

\bibitem{UltraEfficient}
M.~Imani, S.~Gupta, and T.~Rosing, ``Ultra-efficient processing in-memory for
  data intensive applications,'' in \emph{2017 54th ACM/EDAC/IEEE Design
  Automation Conference (DAC)}, 2017, pp. 1--6.

\bibitem{Dadda}
L.~Guckert and E.~E. Swartzlander, ``Dadda multiplier designs using
  memristors,'' in \emph{ICICDT}, 2017, pp. 1--4.

\bibitem{SemiSerialMultiplier}
D.~Radakovits, N.~TaheriNejad, M.~Cai, T.~Delaroche, and S.~Mirabbasi, ``A
  memristive multiplier using semi-serial imply-based adder,'' \emph{IEEE
  Transactions on Circuits and Systems I: Regular Papers}, 2020.

\bibitem{MemristorBasedMultipliers}
L.~Guckert and E.~E. Swartzlander, ``Optimized memristor-based multipliers,''
  \emph{IEEE TCAS-I}, vol.~64, no.~2, pp. 373--385, 2017.

\bibitem{ResistiveComputing}
S.~Shin, K.~Kim, and S.-M. Kang, ``Resistive computing: Memristors-enabled
  signal multiplication,'' \emph{IEEE Transactions on Circuits and Systems I:
  Regular Papers}, vol.~60, no.~5, pp. 1241--1249, 2013.

\bibitem{Ameer}
A.~Haj-Ali, R.~Ben-Hur, N.~Wald, and S.~Kvatinsky, ``Efficient algorithms for
  in-memory fixed point multiplication using {MAGIC},'' in \emph{IEEE
  International Symposium on Circuits and Systems (ISCAS)}, 2018.

\bibitem{IMAGING}
A.~Haj-Ali, R.~Ben-Hur, N.~Wald, R.~Ronen, and S.~Kvatinsky, ``{IMAGING:}
  in-memory algorithms for image processing,'' \emph{IEEE Transactions on
  Circuits and Systems I: Regular Papers}, 2018.

\bibitem{FloatPIM}
M.~Imani, S.~Gupta, Y.~Kim, and T.~Rosing, ``{FloatPIM}: In-memory acceleration
  of deep neural network training with high precision,'' in \emph{Annual
  International Symposium on Computer Architecture}, 2019.

\bibitem{RIME}
Z.~Lu, M.~T. Arafin, and G.~Qu, ``{RIME}: A scalable and energy-efficient
  processing-in-memory architecture for floating-point operations,'' in
  \emph{Asia and South Pacific Design Automation Conference}, 2021.

\bibitem{sunder1995two}
S.~Sunder, F.~El-Guibaly, and A.~Antoniou, ``Two's-complement fast
  serial-parallel multiplier,'' \emph{IEE Proceedings-Circuits, Devices and
  Systems}, vol. 142, no.~1, pp. 41--44, 1995.

\bibitem{richards1955arithmetic}
R.~Richards, \emph{Arithmetic Operations in Digital Computers}, ser. University
  series in higher mathematics.\hskip 1em plus 0.5em minus 0.4em\relax New
  York, 1955.

\bibitem{FSP}
Gnanasekaran, ``A fast serial-parallel binary multiplier,'' \emph{IEEE
  Transactions on Computers}, vol. C-34, no.~8, pp. 741--744, 1985.

\bibitem{XMAGIC}
N.~{Peled}, R.~{Ben-Hur}, R.~{Ronen}, and S.~{Kvatinsky}, ``{X-MAGIC}:
  Enhancing {PIM} using input overwriting capabilities,'' in \emph{VLSI-SoC},
  2020.

\bibitem{SIMPLER}
R.~{Ben-Hur}, R.~{Ronen}, A.~{Haj-Ali}, D.~{Bhattacharjee}, A.~{Eliahu},
  N.~{Peled}, and S.~{Kvatinsky}, ``{SIMPLER MAGIC}: Synthesis and mapping of
  in-memory logic executed in a single row to improve throughput,'' \emph{IEEE
  TCAD}, 2020.

\bibitem{ComputerArithmetic}
M.~Vl{\u{a}}du{\c{t}}iu, \emph{Computer arithmetic: algorithms and hardware
  implementations}.\hskip 1em plus 0.5em minus 0.4em\relax Springer Science \&
  Business Media, 2012.

\end{thebibliography}

\end{document}